\documentclass[nofootinbib,twocolumn,prd,aps,amsmath,superscriptaddress,tightenlines]{revtex4}
\usepackage{graphicx}
\usepackage{bm}
\usepackage{epsfig}

\newif\ifpdf
\ifx\pdfoutput\undefined
\pdffalse 
\else
\pdfoutput=1 
\pdftrue
\fi

\def\OMIT#1{}
\def\mbbar{\bar{m}_B}
\def\lbar{\bar{\Lambda}_{1S}}

\newcommand{\nn}{\nonumber}

\newcommand{\bea}{\begin{eqnarray}}
\newcommand{\eea}{\end{eqnarray}}

\newcommand{\LQCD}{{\Lambda_{\rm QCD}}}

\begin{document}
\ifpdf
\DeclareGraphicsExtensions{.pdf, .jpg}
\else
\DeclareGraphicsExtensions{.eps, .jpg}
\fi


\preprint{ \vbox{\hbox{UCSD/PTH 02-09}\hbox{UTPT 02-06} \hbox{hep-ph/0205039}  }}

\title{\phantom{x}\vspace{0.5cm} 
Reducing theoretical uncertainties in $m_b$ and $\lambda_1$ 
\vspace{0.5cm} }

\author{Christian W.~Bauer}
\affiliation{Department of Physics, University of California at San Diego,
	La Jolla, CA 92093\footnote{Electronic address: bauer@physics.ucsd.edu}}

\author{Michael Trott}
\affiliation{Department of Physics, University of Toronto, 60 St. George St. 
	Toronto, Ontario, M5S-1A7, Canada\footnote{Electronic address: mrtrott@physics.utoronto.ca}}

\date{\today\\ \vspace{1cm} }

\begin{abstract}
We calculate general moments of the lepton energy spectrum in inclusive semileptonic $B \to X_c \ell \bar\nu$ decay. 
Moments which allow the determination of $m^{1S}_b$ and $\lambda_1$ with theoretical uncertainties $\Delta m_b^{1S} \sim 0.04\rm{ 
GeV} $ and $\Delta \lambda_1 \sim 0.05 \rm{GeV}^2$ are presented. The short distance $1S$ mass is used
to extract a mass parameter free of renormalon ambiguities. 
Moments which are insensitive to $m_b$ and  $\lambda_1$ and therefore test the size of the $1/m_b^3$ matrix elements and the validity of the OPE are also presented. Finally, we give an expression for the total branching ratio 
with a lower cut on the lepton energy, which allows one to eliminate a source of
model dependence in current determinations of $|V_{cb}|$ from $B \to X_c \ell \bar\nu$ decay. 

\end{abstract}

\maketitle




Inclusive semileptonic $B$ decays have been examined in recent years using an operator product expansion (OPE)\cite{ope}.
This OPE coincides with the parton 
model in the limit $m_{b} \to \infty $ with nonperturbative corrections suppressed by powers of $ {\LQCD}/{m_{b}}$. There are no nonperturbative corrections
at order ${\LQCD}/{m_{b}}$, and the order $\left({\LQCD}/{m_{b}}\right)^2$ corrections are parameterized by two matrix elements \cite{opecorr1,opecorr2}
\begin{eqnarray}
\lambda_1 &=& \frac{1}{2 m_B} \left<B|\bar{h}_{v}(iD)^2 h_{v}|B\right>\,,\nn\\
\lambda_2 &=& \frac{1}{6 m_B}\left<B\left|\bar{h}_{v}\frac{g}{2}\sigma_{\mu \nu} G^{\mu \nu} h_{v}\right|B\right>\,.
\end{eqnarray}
Here $ h_{v} $ is the heavy quark field in the effective theory. Physically these parameters correspond to matrix 
elements of the kinetic and chromomagnetic operators in the heavy quark effective theory \cite{wisemanohar}.
To second order in $\Lambda_{\rm QCD}/m_b$ the semileptonic decay rate for $B \to X_c \ell \bar\nu$ in the OPE is
\begin{eqnarray}
\Gamma = \frac{G_F^2 m_b^5}{192 \pi^3} |V_{cb}|^2 \left[ f(\rho)\left(1+\frac{\lambda_1}{2m_b^2}\right) + g(\rho)\frac{\lambda_2}{2m_b^2}\right]\,,
\end{eqnarray}
where $ \rho = m_c^2/m_b^2$ and
\begin{eqnarray}
f(\rho) &=& 1-8\rho +8{\rho}^3- {\rho}^4 - 12 {\rho}^2 \log\rho,\\\nn 
g(\rho) &=& -9 + 24 {\rho} - 72 {\rho}^2 + 72 {\rho}^3 - 15 {\rho}^4 -36 {\rho}^2 \log\rho.
\end{eqnarray} 
This inclusive decay can be used to measure the magnitude of the CKM matrix 
element $|V_{cb}|$. The results of this analysis yields a value in good agreement with determinations from exclusive decays \cite{CKMtalk}
\begin{eqnarray}
|V_{cb}| = (40.4 \pm 0.9_{exp} \pm 0.5_{\lambda_1,\bar\Lambda} \pm 0.8_{th}) \times 10^{-3}\,.\end{eqnarray}
The first uncertainty is experimental, the second is due to uncertainties in the
values  of $m_b$ and $\lambda_1$, while the third is from unknown higher order terms in the OPE and unknown perturbative corrections. 
One of the systematic uncertainties is due to modeling the 
spectrum below the required cut on the charged lepton energy spectrum and therefore is also theoretical in nature. For machines running on the $\Upsilon(4S)$ resonance, this experimentally required cut is typically at 0.6 GeV.

The measurement of $|V_{cb}|$ from inclusive decays requires that these decays be adequately described by the OPE formalism. In particular, the possibility of deviations from the OPE predictions due to quark-hadron duality have been raised \cite{Isgurduality}. However, in order to compare the OPE predictions with data, one also has to define how uncertainties from $1/m_b^3$ corrections are estimated. These uncertainties are hard to quantify reliably, 
as the only information on the size of matrix elements of the dimension six operators comes from dimensional analysis. The important question to answer for a reliable determination of $|V_{cb}|$ from inclusive decays is therefore how well the OPE fits the data. Thus, in this paper we define duality violation to be {the difference between the OPE predictions using dimensional scaling of the $1/m_b^3$ matrix elements and the experimental data}.

It is the purpose of this paper to demonstrate how to reduce the theoretical error on the extraction of $|V_{cb}|$,  $m_b$ 
and $\lambda_1$ from semileptonic $B \to X_c \ell \bar\nu$ decay. We eliminate the uncertainty due to modeling the 
spectrum below 0.6 GeV by calculating the dependence on this cut. We demonstrate how to 
improve the precision of the determination of the parameters $m_b$ and $\lambda_1$ using measurements of generalized moments of the lepton energy spectrum. We also calculate several moments which are insensitive on the actual values of $m_b$ and $\lambda_1$ and therefore allow one to test the validity of the OPE and therefore our definition of duality violation.

Past attempts to measure $m_b$ and $\lambda_1$ from the lepton energy spectrum of $B \to X_c \ell \bar\nu$ used the moments presented in \cite{leptonmoment2} 
\begin{eqnarray}\label{R1R2}
R_1 = \frac{\int_{1.5 {\rm GeV}}^{E_{\rm max}} E_\ell \frac{d\Gamma}{d E_\ell}d E_\ell}{\int_{1.5 {\rm GeV}}^{E_{\rm max}} \frac{d\Gamma}{d E_\ell}d E_\ell}\qquad
R_2 = \frac{\int_{1.7 {\rm GeV}}^{E_{\rm max}} \frac{d\Gamma}{d E_\ell}d E_\ell}{\int_{1.5 {\rm GeV}}^{E_{\rm max}} \frac{d\Gamma}{d E_\ell}d E_\ell}\,,
\end{eqnarray}
where $E_{\rm max} = (m_B^2-m_D^2)/2m_B$. 
These moments were calculated to third order in the $\LQCD/m_b$ expansion \cite{leptonmoment3} and to order 
$\alpha_s^2 \beta_0$ \cite{leptonmoment4} in perturbation theory.  The pole masses were expressed in terms of the
heavy meson masses and the parameter $\bar\Lambda$ via the relation
\begin{eqnarray}
m_M = m_Q^{\rm pole} + \bar{\Lambda} - \frac{\lambda_1 + 3 \lambda_2}{2 m_Q} + O\left(\frac{\LQCD^3}{{m_Q}^3}\right)\,.
\end{eqnarray}
It is well known that the pole mass and the parameter $\bar\Lambda$ have a renormalon ambiguity \cite{renormalon}. In every physical 
observable, such as these moments, this ambiguity is canceled by corresponding terms in the perturbative series \cite{luke_renormalon}.  This means, however, 
that a value for the pole mass (or $\bar\Lambda$) only has a meaning if specified at a certain order in perturbation theory. 
While this is not a problem in principle, it is much more convenient to work with a short distance mass which does not exhibit such a
renormalon ambiguity. Using a short distance mass instead of the pole mass also tends to give a perturbative expansion which converges
more rapidly.

In this paper we calculate the general moments of the lepton energy spectrum\footnote{In general, non-integer moments can lead to new cuts in the complex plane which have to be dealt with carefully. This, however, is not a problem for this paper.}
\begin{eqnarray}\label{momentdefinition}
R [n,E_{\ell_1},m,E_{\ell_2}] = \frac{\int_{E_{\ell_1}}^{E_\ell^{\rm max}} E_\ell^n \frac{d\Gamma}{d E_\ell}d E_\ell}{\int_{E_{\ell_2}}^{E_\ell^{\rm max}} E_{\ell}^m 
\frac{d\Gamma}{d E_\ell}d E_\ell}\,,
\end{eqnarray}
to order $(\LQCD/m_b)^3$ and $\alpha_s^2 \beta_0$ using the short distance 1S mass and systematically examine what information can be
obtained from these moments.  In particular, we identify sets of moments which allow  $m_b$ and $\lambda_1$ to be measured with
uncertainties considerably smaller than $R_1$ and $R_2$. We also find moments which are insensitive to the values of $m_b$ and $\lambda_1$
and therefore directly test  the reliability of the theoretical error assigned to the extraction of $m_b$ and $\lambda_1$ and the
theoretical treatment of inclusive $B$ decay using an OPE. 

The moments are calculated from the lepton energy spectrum, which has been calculated to order $(\Lambda/m_b)^3$ \cite{leptonmoment3}, 
and to order $\alpha_s^2 \beta_o$ perturbatively \cite{leptonmoment4,alphaspec}. The calculated moments depend on the matrix elements of the
higher dimensional operators and the pole masses of the $b$ and $c$ quarks. To avoid the renormalon ambiguity, we rewrite the expressions for 
the moments in terms of the 1S mass, which is related to the b quark pole mass through the relation 
\cite{renorupsilon,upmassmanohar}
\begin{eqnarray}\label{masstrans}
\frac{m_b^{\rm 1S}}{ m_b^{\rm pole}} &=& 1 - \frac{(\alpha_s C_F)^2}{8} \Big\{ 1\epsilon  +\nn\\
&&\left.  \qquad\frac{\alpha_s}{\pi} \left[ \left(\ell+\frac{11}{6} \right) \beta_0 -4 \right] 
\epsilon^2 +
 \ldots \right\},
\end{eqnarray}
where $ \ell = \log[ \mu / (m_{b} \alpha_{s} C_{F} ) ] $ and $C_F = 4/3$. The dependence on the pole mass of the charm quark is eliminated through the relation
\begin{eqnarray}\label{charmrelation}
m_b^{\rm pole} - m_c^{\rm pole} &=& \bar{m}_B - \bar{m}_D + \lambda_1 \frac{\bar m_D-\bar m_B}{2\bar{m}_B \bar m_D} 
\\\nn 
&&\hspace{-1cm}+ \left(\tau_1 + \tau_3 - \rho_1 + \lambda_{1} \, \bar\Lambda_{1S} \right)\frac{\bar m_D^2-\bar m_B^2}{4\bar{m}_B^2 \bar m_D^2}\,,
\end{eqnarray}
where $\rho_1$ and $\tau_i$ are matrix elements of local dimension six operators \cite{leptonmoment3}.
In this relation  we use the fact that 
$\bar{m}_B-m_b^{1S} \sim \Lambda_{\rm QCD}$ by expanding in the parameter $\bar{\Lambda}_{1S}$ 
\begin{eqnarray}
\bar{\Lambda}_{1S} \equiv \bar{m}_B - m_b^{1S}\,.
\end{eqnarray} 
It is important to note that in contrast to $\bar\Lambda$, which relates the pole mass 
to the meson masses, the parameter $\bar{\Lambda}_{1S}$ is not an infrared sensitive quantity. 
Using $m_b^{1S}$ necessitates introducing a modified perturbative expansion in order to ensure the cancellation of renormalon
ambiguities \cite{renorupsilon}. When calculating in the 1S mass scheme the order ${\alpha_{s}}^n$ perturbative
corrections coming from the mass transformation Eq.~(\ref{masstrans}) are counted using the parameter $\epsilon^{n-1}$ while $\alpha_s^n$ corrections in the decay rate 
are counted 
as $\epsilon^{n}$.  The parameter $\epsilon$ determines the order in the modified perturbative expansion.

The lepton energy spectrum has been measured by the CLEO collaboration using single and double tagged charged lepton data 
samples \cite{cleoprl,wang_thesis} . The double tagged data uses the charge correlation between primary and secondary leptons to eliminate the background from secondary 
leptons. While this type of analysis gives the smallest experimental uncertainties, it does require a correct identification of the primary 
lepton. For this reason a cut on the lepton energy at $E_\ell = 1.4\, {\rm GeV}$ was employed by CLEO \cite{cleoprl}. 
The double tagged data sample does not require such a cut, and one only needs a lower cut on the lepton energy at $E_\ell = 0.6$ GeV to eliminate fake pion signals. 
However, this increases the uncertainties in the moments by about a factor of three \cite{wang_thesis}. With the large data samples available 
to the $B$ factories it should be possible to use single tagged data samples to measure the
moments with the required precision. For each search of optimal moments we therefore present results relevant for the two analysis techniques by restricting the lepton energy to lie in the conservative regions above $800\, {\rm MeV}$ and above  $1.5\, {\rm GeV}$, respectively.

To compare with experimental data at the required precision, we also calculate the corrections due to the electroweak electron radiative corrections~\cite{atwood},
the leading order corrections proportional to $|{V_{ub}}/{V_{cb}}|^2 $, and the corrections due to a boost from the rest frame of the $B$ meson to the lab frame. 
For the  boost corrections we assume a mono-energetic B meson of energy $ E = M_{\Upsilon}^{(4S)}/2$ appropriate for the CLEO analysis. Boost corrections for the 
asymmetric $B$ factories are not considered here and should be taken into account in the experimental analysis. In this case, the boost corrections calculated here should not be included. 

Before we present the results of the general moments search, we can use our result to obtain the total branching ratio with a cut on the lepton energy. 
As explained earlier, such a cut is required experimentally to avoid fake lepton backgrounds. Here we present the branching ratio as a function of the 
lower cut on $E_\ell$. Since the dependence on the cut is quite complicated, we will not present an analytic equation, but rather give an approximate formula, 
which deviates from the exact result by less than 1\% for $0< E_{\rm cut}<0.8 \, {\rm GeV}$
\begin{widetext}
\begin{eqnarray}
\int_{E_{\rm cut}}^{E_\ell^{\rm max}} \frac{d\Gamma}{d E_\ell} &=& 
\frac{G_F^2 |V_{cb}|^2 \bar m_B^5}{192 \pi^3} 
\left[\left(0.356+0.12\hat E_{\rm cut} - 2.4 \hat E_{\rm cut}^2\right) 
- \left(0.617 - 0.31 \hat E_{\rm cut}\right) \frac{\lbar}{\bar m_B}
\right.\nonumber\\
&&  \hspace{2.3cm}\left.
- \left(0.35 - 0.14 \hat E_{\rm cut}\right) \frac{\lbar^2}{\bar m_B^2} 
- \left(1.49 - 0.24 \hat E_{\rm cut}\right) \frac{\lambda_1}{\bar m_B^2}
- \left(0.034 - 0.02 \hat E_{\rm cut}\right) \epsilon
\right.\\
&&  \hspace{2.3cm}\left.
+ \left(0.04 - 0.02 \hat E_{\rm cut}\right) \epsilon \frac{\lbar}{\bar m_B}
-  \left(0.01 - 0.01 \hat E_{\rm cut}\right) \epsilon^2_{\rm BLM}
\right]\,.\nn
\end{eqnarray}
\end{widetext}
Here $\hat E_{\rm cut} = E_{\rm cut}/\bar m_B$ and we have used $\alpha_s = 0.22$, $\beta_0 = 25/3$ and $\lambda_2 = 0.12\, {\rm GeV}^2$.  The dependence of the uncertainties in this expression from perturbative and $1/m_b^3$ corrections is negligible over the range of  $E_{\rm cut}$ considered, and can thus be taken as constant. 

For the measurement of $m_b$ and $\lambda_1$, we compare the resulting uncertainties on the extracted parameters with those obtained using the moments $R_1$ and $R_2$ defined in (\ref{R1R2}). To facilitate this comparison, we present these moments in terms of the 1S mass
\begin{widetext}
\begin{eqnarray}
R_1 &=& R[1,1.5,0,1.5]= \label{R1}
1.8056\Big[1 
-0.17 \frac{\lbar}{\mbbar} 
-0.20 \frac{\lbar^2}{\mbbar^2} - 1.37 \frac{\lambda_1}{\mbbar^2} - 2.19 \frac{\lambda_2}{\mbbar^2}
-0.2 \frac{\lbar^3}{\mbbar^3}
-3.5 \frac{\lbar \lambda_1}{\mbbar^3} 
 \\
&\,& \hspace{.1cm} 
- 3.8 \frac{\lbar \lambda_2}{\mbbar^3}  
- 4.2 \frac{\rho_1}{\mbbar^3} - 
0.7 \frac{\rho_2}{\mbbar^3} 
- 1.8 \frac{\tau_1}{\mbbar^3} 
- 2.5 \frac{\tau_2}{\mbbar^3}
 -1.7 \frac{\tau_3}{\mbbar^3} 
-2.2 \frac{\tau_4}{\mbbar^3}
+\epsilon\,\left(0.0005 - 0.0005 \frac{\lbar}{\mbbar}\right) 
  \nn\\
&\,& \hspace{.1cm}\left. 
+ \epsilon^2_{\rm BLM} 0.0014 
+ {\left|\frac{V_{ub}}{V_{cb}}\right|}^2 \left(0.74 - 5.7 \frac{\lbar}{\mbbar}\right) 
- \left(0.0023 - 0.002 \frac{\lbar}{\mbbar}\right) 
+ \left(0.0034 + 0.001 \frac{\lbar}{\mbbar}\right) \right]\,, \nn\\
R_2 &=& R[0,1.7,0,1.5] = \label{R2}
0.6578\Big[1 
- 0.48 \frac{\lbar}{\mbbar} 
- 1.03 \frac{\lbar^2}{\mbbar^2} 
- 2.76 \frac{\lambda_1}{\mbbar^2} 
- 7.52 \frac{\lambda_2}{\mbbar^2} 
- 2.30 \frac{\lbar^3}{\mbbar^3}
- 12.1 \frac{\lbar \lambda_1}{\mbbar^3} 
 \\&\,& \hspace{.1cm}
- 26.6 \frac{\lbar \lambda_2}{\mbbar^3}  
- 2.7 \frac{\rho_1}{\mbbar^3} + 
3.5 \frac{\rho_2}{\mbbar^3} 
- 4.5 \frac{\tau_1}{\mbbar^3} 
- 2.2 \frac{\tau_2}{\mbbar^3} 
- 6.2 \frac{\tau_3}{\mbbar^3} 
- 7.5 \frac{\tau_4}{\mbbar^3}
+\epsilon\, \left(0.0010 - 0.002 \frac{\lbar}{\mbbar}\right) 
\nn \\&\,& \hspace{.1cm} \left. 
+ \epsilon^2_{\rm BLM} 0.0031 
+ {|\frac{V_{ub}}{V_{cb}}|}^2 \left(1.33 - 5.8 \frac{\lbar}{\mbbar}\right)  
- \left(0.0113 + 0.008 \frac{\lbar}{\mbbar}\right) 
+ \left(0.0031 + 0.005 \frac{\lbar}{\mbbar}\right)\right]\,. \nn
\end{eqnarray}
\end{widetext}
The last two brackets in each moment are the electroweak and boost correction, respectively.
Using the central values and statistical correlation matrix $V$ measured by the CLEO collaboration 
\begin{eqnarray}
R_1 &=&  1.7831\,, \quad R_2 = 0.6159\,, \nn\\
V(R_1,R_2) &=& 
\left(\begin{array}{ll}
3.8 \times 10^{-6} & 6.0 \times 10^{-6}\\
6.0 \times 10^{-6} & 1.7 \times 10^{-5}
\end{array}\right)\,,\nn
\end{eqnarray}
together with $|V_{ub}/V_{cb}| = 0.1\pm 0.03$ and $\alpha_s = 0.22$, we obtain 
\begin{eqnarray}
\lbar \!\!\!&=&\!\!\! \left[0.47 \, \pm 0.10_{exp} \pm 0.02_{V_{ub}} \pm 0.02_\epsilon \pm 0.07_{m^3}\right] {\rm GeV}\nn\\ 
\lambda_1 \!\!\!&=&\!\!\! \left[-0.16 \pm 0.11_{exp} \pm 0.02_{V_{ub}} \pm 0.02_\epsilon \pm 0.07_{m^3}\right] {\rm GeV}^2\,,\nn
\end{eqnarray}
where the perturbative uncertainties are estimated by varying $m_b/2 < \mu < 2m_b$ and using the ${\cal O}(\epsilon^2 \beta_0)$ term as a crosscheck. The uncertainty due to the unknown size of matrix elements of $1/m_b^3$ operators is estimated by using the method of \cite{leptonmoment3,mypapers} to ensure a valid comparison of
our suggested observables with $R_1$ and $R_2$: We impose the relation due to the vector, pseuodoscalar mass splitting at third order \cite{leptonmoment3}
\begin{eqnarray}
\rho_2 - \tau_2 - \tau_4 &=& \nn\\
&&\hspace{-1.5cm}
\frac{\kappa\,  m_B^2 \Delta m_B(m_D + \lbar) - m_D^2 \Delta m_D (m_B + \lbar)}{m_B + \lbar - \kappa\,(m_D + \lbar)}\,,\nn
\end{eqnarray}
and use the result from the  vacuum saturation approximation which predicts a positive value of $\rho_1$ \cite{vacuumsat}. The unknown 
matrix elements are then randomly varied between $(\pm 500\,{\rm MeV})^3$. The random values are drawn from a flat distribution, since there is no known  preferred value of the individual matrix elements. 

We would like to make a few comments about the expressions for $R_1$ (\ref{R1}) and $R_2$ (\ref{R2}). First, the expressions differ from the ones presented in \cite{leptonmoment2} in their perturbative terms and the terms containing $\lambda_1$. This is due to the use of a short distance mass in the present paper, as supposed to the pole mas used in \cite{leptonmoment2}. Furthermore, in both moments the ${\cal O}(\epsilon^2 \beta_0)$ term in the perturbative series is larger than the order $\epsilon$ term. This, however, is due to a cancellation of the ${\cal O}(\epsilon)$ terms in the ratios considered, and does not indicate a poorly behaved perturbation series. In both the numerator and the denominator of these expressions the perturbation series is well behaved.

Translating the obtained value of $\lbar$ into a value of the $b$ quark mass we find $m_b^{1S} = 4.84 \pm 0.13$, adding the errors in quadrature.
This result is in agreement with both sum rule and other moment analysis extractions\cite{hoang}, with comparable uncertainties. Using the definition of the general moments given in (\ref{momentdefinition}), one can search for combinations of moments which allow for a determination of $m_b^{1S}$ and $\lambda_1$ with smaller uncertainties. For certain values of $n$, $m$, $E_{\ell_1}$ and $E_{\ell_2}$, the OPE results can not be trusted any more since the convergence of the OPE breaks down. 
Using dimensional analysis to obtain the size of the matrix elements and requiring that the $1/m_b^2$ and $1/m_b^3$ 
contributions do not exceed 10\% and 3\% of the leading order result, respectively, we restrict ourself to the parameter space
\begin{eqnarray}
{m}<3\,,\, n<3\,, \quad 0.8\, {\rm GeV} < E_{\ell_i}^{\rm cut}< 1.8\, {\rm GeV}\, 
\end{eqnarray}
in our search to ensure a well behaved OPE.

\begin{figure}
\includegraphics[width=7.5cm]{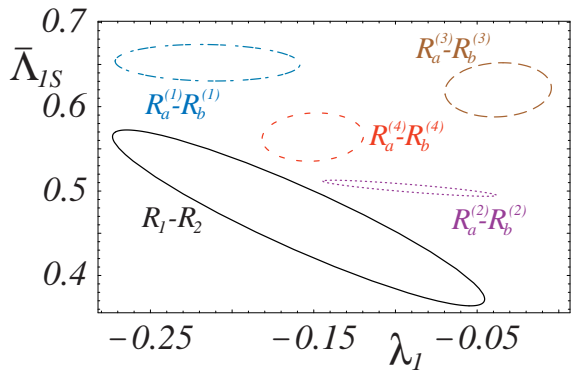}
\caption{Comparison of the error ellipses due to the $1/m_b^3$ uncertainties for the various sets of moments. Only the central value of the black, solid ellipse has meaning, for all the other ellipses only their relative size is important.\label{fig1}}
\end{figure}

By simultaneously minimizing the size of third order corrections and maximizing 
the linear independence of the moments we find sets of observables $R_a$ and $R_b$ which minimize the theoretical uncertainty
in extracting  $\lbar$ and $\lambda_1$. We do not try to minimize the size of the perturbative corrections. Since the full two loop correction to the lepton spectrum is not available, it would be meaningless to minimize the size if the $\epsilon^2 \beta_0$ piece.
Consider
\begin{widetext}
\begin{eqnarray}
R^{(1)}_{a} &=& R[1.4,1.3,1,1]\hspace{0.1cm}  = \hspace{0.1cm}
1.0441\Big[1 
- 0.20 \frac{\lbar}{\mbbar} 
- 0.33 \frac{\lbar^2}{\mbbar^2} 
- 1.53 \frac{\lambda_1}{\mbbar^2} 
- 3.66 \frac{\lambda_2}{\mbbar^2}
- 0.6 \frac{\lbar^3}{\mbbar^3}
- 4.9 \frac{\lbar \lambda_1}{\mbbar^3} 
 \\
&\,& \hspace{.1cm} 
- 8.3 \frac{\lbar \lambda_2}{\mbbar^3}  
- 3.4 \frac{\rho_1}{\mbbar^3} 
+ 1.3 \frac{\rho_2}{\mbbar^3} 
- 2.4 \frac{\tau_1}{\mbbar^3} 
- 1.4 \frac{\tau_2}{\mbbar^3} 
- 3.2 \frac{\tau_3}{\mbbar^3} 
- 3.7 \frac{\tau_4}{\mbbar^3}
+\epsilon\, \left(0.0016 + 0.001 \frac{\lbar}{\mbbar}\right) 
 \nn\\&\,& \hspace{.1cm} \left.
+ \epsilon^2_{\rm BLM} 0.0019
+ {\left|\frac{V_{ub}}{V_{cb}}\right|}^2 \left(0.81 - 5.2 \frac{\lbar}{\mbbar}\right) 
- \left(0.0088 + 0.003\frac{\lbar}{\mbbar}\right) 
+ \left(0.0019 - 0.0001 \frac{\lbar}{\mbbar}\right) \right] \nn\\
R^{(1)}_{b} &=& R[1.7,1.4,1.2,0.8]\hspace{0.1cm}  = \hspace{0.1cm}
0.9822\Big[1 
- 0.34 \frac{\lbar}{\mbbar} 
- 0.54 \frac{\lbar^2}{\mbbar^2} 
- 2.52 \frac{\lambda_1}{\mbbar^2} 
- 6.02 \frac{\lambda_2}{\mbbar^2}
- 1.0 \frac{\lbar^3}{\mbbar^3}
- 8.0 \frac{\lbar \lambda_1}{\mbbar^3} 
 \\
&\,& \hspace{.1cm} 
- 13.7 \frac{\lbar \lambda_2}{\mbbar^3}  
- 5.7 \frac{\rho_1}{\mbbar^3} 
+ 2.1 \frac{\rho_2}{\mbbar^3} 
- 4.0 \frac{\tau_1}{\mbbar^3} 
- 2.3 \frac{\tau_2}{\mbbar^3} 
- 5.2 \frac{\tau_3}{\mbbar^3} 
- 6.0 \frac{\tau_4}{\mbbar^3}
+\epsilon\, \left(0.0027 + 0.002 \frac{\lbar}{\mbbar}\right) 
 \nn\\&\,& \hspace{.1cm} \left.
+ \epsilon^2_{\rm BLM} 0.0032
+ {\left|\frac{V_{ub}}{V_{cb}}\right|}^2 \left(1.33 - 8.7 \frac{\lbar}{\mbbar}\right) 
- \left(0.0146 + 0.005\frac{\lbar}{\mbbar}\right) 
+ \left(0.0032 + 0.0002 \frac{\lbar}{\mbbar}\right) \right] \,.\nn
\end{eqnarray}
\end{widetext}
For this moment, the only available data is CLEO's double tagged lepton spectrum, and the uncertainties are too large to be useful for this analysis. Thus, to illustrate how well the parameters $\lbar$ and $\lambda_1$ can be extracted from these moments we use the hypothetical data $R^{(1)}_{a} = 1.0082$ and $R^{(1)}_{b} = 0.9266$. Using again $|V_{ub}/V_{cb}|^2 = 0.1 \pm 0.03$ and $\alpha_s = 0.22$, this leads to 
$\lbar = [0.59 \pm 0.02_{V_{ub}} \pm 0.03_\epsilon \pm 0.01_{m^3}] \,{\rm GeV}$, and 
$\lambda_1 = [-0.16 \pm 0.04_{V_{ub}} \pm 0.02_\epsilon \pm 0.03_{m^3}] \,{\rm GeV^2}$. Adding all errors in quadrature, this leads to a theoretical error on $m_b^{1S}$ of $\pm 40$ MeV, with the largest error from the perturbative uncertainties. To further reduce the error, a full two loop calculation for the lepton energy spectrum would be required. The largest uncertainty on $\lambda_1$ is due to the error in $|V_{ub}|$, for which we have assumed a very conservative error. Future measurements of $|V_{ub}|$ should lower this uncertainty considerably \cite{Vub}.
In Fig.~\ref{fig1} we compare the resulting 68\% confidence level ellipses in the $\lbar-\lambda_1$ plane with the one obtained from $R_1$ and $R_2$. To estimate how the experimental error from these new moments compares with that from the previous extraction, we scale the measured correlation matrix for $R_1$ and $R_2$ by the central values to keep the percentage error equal.  This leads to slightly increased experimental uncertainties compared with those using $R_1$ and $R_2$. Thus, to minimize the overall uncertainty, the  new moments have to be measured with smaller uncertainties than $R_1$ and $R_2$  \cite{wang_thesis}. Considering that this measurement was based on only $2\, fb^{-1}$ of data, this seems feasible. 

Next we turn our attention to moments that are insensitive to the values of $m_b$ and $\lambda_1$ and are therefore well suited to test the underlying assumptions of the OPE and thus for duality violations. Consider 
\begin{widetext}
\begin{eqnarray}
D_1 &=& R[0.2,1.3,1,1]\hspace{0.1cm}  = \hspace{0.1cm}
0.5449\Big[1 
+ 0.02 \frac{\lbar}{\mbbar} 
- 0.08 \frac{\lbar^2}{\mbbar^2} 
+ 0.17 \frac{\lambda_1}{\mbbar^2} 
- 0.60 \frac{\lambda_2}{\mbbar^2} 
- 0.3 \frac{\lbar^3}{\mbbar^3}
- 0.7 \frac{\lbar \lambda_1}{\mbbar^3} 
\\&\,& \hspace{.1cm} 
- 3.3 \frac{\lbar \lambda_2}{\mbbar^3}  
+ 1.5 \frac{\rho_1}{\mbbar^3} 
+ 1.7 \frac{\rho_2}{\mbbar^3} 
- 0.1 \frac{\tau_1}{\mbbar^3} 
+ 1.3 \frac{\tau_2}{\mbbar^3}  
- 0.7 \frac{\tau_3}{\mbbar^3} 
- 0.6 \frac{\tau_4}{\mbbar^3}
+\epsilon\, \left(0.0009 + 0.002 \frac{\lbar}{\mbbar}\right) 
\nn\\&\,& \hspace{.1cm} \left.
+ \epsilon^2_{\rm BLM} 0.0002 
- {\left|\frac{V_{ub}}{V_{cb}}\right|}^2 \left(0.12 - 1.9 \frac{\lbar}{\mbbar}\right) 
- \left(0.0047 + 0.008\frac{\lbar}{\mbbar}\right) 
- \left(0.0018 + 0.0003 \frac{\lbar}{\mbbar}\right) \right]\,. \nn
\end{eqnarray}
\end{widetext}
Using the same input parameters as before, and very conservative errors for the two HQET parameters $\lbar= 0.6 \pm 0.2$, $\lambda_1 = -0.3 \pm 0.3$, this moments is predicted to be 
\begin{eqnarray}
D_1 &\!\!=\!\!& 0.5400 \left[1 \pm 0.0010_{\lambda_1} \pm 0.0003_{\lbar} \pm 0.0006_{V_{ub}}\right.\nn\\
&&\left.\hspace{1.4cm}
\pm 0.0002_\epsilon \pm 0.0009_{m^3}\right] \,.\nn
\end{eqnarray}
Thus, even with little information on the size of $\lbar$ and $\lambda_1$, the numerical value of this moment is predicted to better than 1\% in the OPE, given that the method to estimate the $1/m_b^3$ uncertainties is reliable. A measurement of this moment can therefore directly test these underlying assumptions, and any deviation from the OPE, even if below the \% level, should be detectable. In the appendix we give several other moments which have similar theoretical uncertainties. While each of these moments individually tests the accuracy  of the OPE predictions, it is best to measure several of these moments. This will give important information on the validity of using the OPE for other inclusive measurements, such as the extraction of $|V_{cb}|$. 

We have presented observables that allow one to improve on the measurement of the parameters $m_b^{\rm 1S}$ and $\lambda_1$. The theoretical error in the extraction of both observables is improved, with the largest remaining uncertainty being from the perturbative expansion. Thus, to further reduce the error, a full two loop calculation of the lepton energy spectrum would be required. We have further presented moments which are insensitive to the size of these two parameters and therefore allow to directly test the underlying assumptions of the OPE. Both measurements are important to reduce and gain confidence in the theoretical uncertainties present in the inclusive determination of $|V_{cb}|$. The method of searching a general set of moments could also be extended to different decay distributions, such as the hadronic invariant mass distribution. 

We would like to thank Zoltan Ligeti and Oliver Buchm{\"u}ller for comments on the manuscript. M.T. would like to thank Craig Burrell, Michael Luke and Alex Williamson for discussions. This work was supported by the DOE under grant DOE-FG03-97ER40546. 

\appendix*
\begin{widetext}
\section{Additional Moments}
In this appendix we present several additional moments which either minimize the theoretical uncertainties on the nonperturbative parameters $\lbar$ and $\lambda_1$ or which are insensitive to these parameters. For each of these categories we give moments subject to the constraint $E_\ell > 1.5$ GeV and moments in which we consider the full range of $E_\ell$. 

\subsection{Moments to extract $\lbar$ and $\lambda_1$ with no restrictions on $E_\ell$ }

\underline{\bf Moments $R[1.4,1.3,0.8,0.9]$ and $R[1.6,1.4,0.9,0.8]$}
\begin{eqnarray}
R^{(2)}_{a} &=& R[1.4,1.3,0.8,0.9]\hspace{0.1cm}  = \hspace{0.1cm}
1.1072\Big[1 
- 0.26 \frac{\lbar}{\mbbar} 
- 0.41 \frac{\lbar^2}{\mbbar^2} 
- 2.00 \frac{\lambda_1}{\mbbar^2} 
- 4.74 \frac{\lambda_2}{\mbbar^2}
- 0.7 \frac{\lbar^3}{\mbbar^3}
- 6.1 \frac{\lbar \lambda_1}{\mbbar^3} 
 \\
&\,& \hspace{.1cm} 
- 10.0 \frac{\lbar \lambda_2}{\mbbar^3}  
- 4.5 \frac{\rho_1}{\mbbar^3} 
+ 1.6 \frac{\rho_2}{\mbbar^3} 
- 3.2 \frac{\tau_1}{\mbbar^3} 
- 1.9 \frac{\tau_2}{\mbbar^3} 
- 4.1 \frac{\tau_3}{\mbbar^3} 
- 4.7 \frac{\tau_4}{\mbbar^3}
+\epsilon\, \left(0.0022 + 0.001 \frac{\lbar}{\mbbar}\right) 
 \nn\\&\,& \hspace{.1cm} \left.
+ \epsilon^2_{\rm BLM} 0.0026
+ {\left|\frac{V_{ub}}{V_{cb}}\right|}^2 \left(1.05 - 7.0 \frac{\lbar}{\mbbar}\right) 
- \left(0.0117 + 0.003\frac{\lbar}{\mbbar}\right) 
+ \left(0.0025 - 0.0001 \frac{\lbar}{\mbbar}\right) \right] \nn\\
R^{(2)}_{b} &=& R[1.6,1.4,0.9,0.8]\hspace{0.1cm}  = \hspace{0.1cm}
1.0615\Big[1 
- 0.38 \frac{\lbar}{\mbbar} 
- 0.59 \frac{\lbar^2}{\mbbar^2} 
- 2.84 \frac{\lambda_1}{\mbbar^2} 
- 6.76 \frac{\lambda_2}{\mbbar^2}
- 1.0 \frac{\lbar^3}{\mbbar^3}
- 8.6 \frac{\lbar \lambda_1}{\mbbar^3} 
 \\
&\,& \hspace{.1cm} 
- 14.5 \frac{\lbar \lambda_2}{\mbbar^3}  
- 6.4 \frac{\rho_1}{\mbbar^3} 
+ 2.4 \frac{\rho_2}{\mbbar^3} 
- 4.5 \frac{\tau_1}{\mbbar^3} 
- 2.6 \frac{\tau_2}{\mbbar^3} 
- 5.9 \frac{\tau_3}{\mbbar^3} 
- 6.8 \frac{\tau_4}{\mbbar^3}
+\epsilon\, \left(0.0030 + 0.002 \frac{\lbar}{\mbbar}\right) 
 \nn\\&\,& \hspace{.1cm} \left.
+ \epsilon^2_{\rm BLM} 0.0036
+ {\left|\frac{V_{ub}}{V_{cb}}\right|}^2 \left(1.50 - 9.9 \frac{\lbar}{\mbbar}\right) 
- \left(0.0163 + 0.004\frac{\lbar}{\mbbar}\right) 
+ \left(0.0036 - 0.0002 \frac{\lbar}{\mbbar}\right) \right] \,.\nn
\end{eqnarray}
Using the hypothetical data $R^{(2)}_{a} = 0.9096$ and $R^{(2)}_{b} = 1.5666$ we find
\begin{eqnarray}
\lbar = (0.62 \, \pm 0.01_{V_{ub}} \pm 0.04_\epsilon \pm 0.01_{m^3}) \, {\rm GeV}\,, \quad 
\lambda_1 = (-0.19 \,\pm 0.01_{V_{ub}}\pm 0.06_\epsilon \pm 0.03_{m^3}) \, {\rm GeV}^2\,.
\end{eqnarray}

\subsection{Moments to extract $\lbar$ and $\lambda_1$, restricted to $E_\ell > 1.5\, {\rm GeV}$}

\underline{\bf Moments $R[0.7,1.7,2,1.5]$ and $R[0.9,1.6,0,1.7]$}
\begin{eqnarray}
R^{(3)}_{a} &=& R[0.7,1.7,2,1.5]\hspace{0.1cm}  = \hspace{0.1cm}
0.3141\Big[1 
- 0.23 \frac{\lbar}{\mbbar} 
-0.83 \frac{\lbar^2}{\mbbar^2} 
- 0.80 \frac{\lambda_1}{\mbbar^2} 
- 4.26 \frac{\lambda_2}{\mbbar^2} 
-2.3 \frac{\lbar^3}{\mbbar^3}
-8.4 \frac{\lbar \lambda_1}{\mbbar^3} 
\\&\,& \hspace{.1cm} 
- 23.2 \frac{\lbar \lambda_2}{\mbbar^3}  
+ 3.7 \frac{\rho_1}{\mbbar^3} 
+ 4.3 \frac{\rho_2}{\mbbar^3} 
- 1.8 \frac{\tau_1}{\mbbar^3} 
+ 1.2 \frac{\tau_2}{\mbbar^3}
- 3.6 \frac{\tau_3}{\mbbar^3} 
-4.3 \frac{\tau_4}{\mbbar^3}
+\epsilon\, \left(0.0003 - 0.001 \frac{\lbar}{\mbbar}\right) 
  \nn\\&\,& \hspace{.1cm} \left.
+ \epsilon^2_{\rm BLM} 0.0011 
+ {\left|\frac{V_{ub}}{V_{cb}}\right|}^2 \left(0.22 + 3.7 \frac{\lbar}{\mbbar}\right) 
- \left(0.0076 + 0.014\frac{\lbar}{\mbbar}\right) 
- \left(0.0016 - 0.006 \frac{\lbar}{\mbbar}\right) \right] \nn\\
R^{(3)}_{b} &=& R[0.9,1.6,0,1.7]\hspace{0.1cm}  = \hspace{0.1cm}
2.2041\Big[1 
+ 0.14 \frac{\lbar}{\mbbar} 
+ 0.55 \frac{\lbar^2}{\mbbar^2} 
+0.40 \frac{\lambda_1}{\mbbar^2} 
+ 2.58 \frac{\lambda_2}{\mbbar^2} 
+ 1.7 \frac{\lbar^3}{\mbbar^3}
+4.9 \frac{\lbar \lambda_1}{\mbbar^3} 
\\&\,& \hspace{.1cm} 
+15.6 \frac{\lbar \lambda_2}{\mbbar^3}  
- 2.7 \frac{\rho_1}{\mbbar^3}  
- 2.9 \frac{\rho_2}{\mbbar^3} 
+ 1.0 \frac{\tau_1}{\mbbar^3} 
- 0.9 \frac{\tau_2}{\mbbar^3}  
+ 2.2 \frac{\tau_3}{\mbbar^3} 
+ 2.6 \frac{\tau_4}{\mbbar^3}
-\epsilon\, \left(0.0001 -0.001 \frac{\lbar}{\mbbar}\right) 
\nn\\&\,& \hspace{.1cm} \left.
- \epsilon^2_{\rm BLM} 0.0006
- {\left|\frac{V_{ub}}{V_{cb}}\right|}^2 \left(0.11 + 2.3 \frac{\lbar}{\mbbar}\right) 
+ \left(0.0044 + 0.010\frac{\lbar}{\mbbar}\right) 
+ \left(0.0013 - 0.002 \frac{\lbar}{\mbbar}\right) \right] \nn
\end{eqnarray}
Using the hypothetical data $R^{(3)}_{a} = 0.2955$ and $R^{(3)}_{b} = 2.2908$ we find
\begin{eqnarray}
\lbar = \left[0.64 \pm 0.01_{V_{ub}} \pm 0.01_\epsilon \pm 0.02_{m^3}\right] {\rm GeV}\,, \quad 
\lambda_1 = \left[-0.18 \pm 0.05_{V_{ub}} \pm 0.04_\epsilon \pm 0.02_{m^3}\right] {\rm GeV^2}\,.
\end{eqnarray}

\underline{\bf Moments $R[0.8,1.6,0,1.7]$ and $R[2.5,1.6,2.9,1.5]$}
\begin{eqnarray}
R^{(4)}_{a} &=& R[0.8,1.6,0,1.7]\hspace{0.1cm}  = \hspace{0.1cm}
2.0712\Big[1 
+ 0.16 \frac{\lbar}{\mbbar} 
+ 0.57 \frac{\lbar^2}{\mbbar^2} 
+ 0.55 \frac{\lambda_1}{\mbbar^2} 
+ 2.79 \frac{\lambda_2}{\mbbar^2} 
+ 1.7 \frac{\lbar^3}{\mbbar^3}
+ 5.4 \frac{\lbar \lambda_1}{\mbbar^3} 
\\&\,& \hspace{.1cm} 
+16.0 \frac{\lbar \lambda_2}{\mbbar^3}  
-2.1 \frac{\rho_1}{\mbbar^3} 
-2.7 \frac{\rho_2}{\mbbar^3} 
+1.2 \frac{\tau_1}{\mbbar^3} 
-0.6 \frac{\tau_2}{\mbbar^3}
+2.3 \frac{\tau_3}{\mbbar^3} 
+2.8 \frac{\tau_4}{\mbbar^3}
-\epsilon\, \left(0.0001 - 0.001 \frac{\lbar}{\mbbar}\right) 
  \nn\\&\,& \hspace{.1cm} \left.
- \epsilon^2_{\rm BLM} 0.0008 
- {\left|\frac{V_{ub}}{V_{cb}}\right|}^2 \left(0.19 + 1.7 \frac{\lbar}{\mbbar}\right) 
+ \left(0.0046 + 0.010\frac{\lbar}{\mbbar}\right) 
+ \left(0.0008 - 0.003 \frac{\lbar}{\mbbar}\right) \right] \nn\\
R^{(4)}_{b} &=& R[2.5,1.6,2.9,1.5]\hspace{0.1cm}  = \hspace{0.1cm}
0.6933\Big[1 
-0.09 \frac{\lbar}{\mbbar} 
-0.30 \frac{\lbar^2}{\mbbar^2} 
-0.40 \frac{\lambda_1}{\mbbar^2} 
-1.65 \frac{\lambda_2}{\mbbar^2} 
-0.8 \frac{\lbar^3}{\mbbar^3}
-3.6 \frac{\lbar \lambda_1}{\mbbar^3} 
\\&\,& \hspace{.1cm} 
-8.7 \frac{\lbar \lambda_2}{\mbbar^3}  
+0.9 \frac{\rho_1}{\mbbar^3}  
+1.4 \frac{\rho_2}{\mbbar^3} 
-0.8 \frac{\tau_1}{\mbbar^3} 
+0.2 \frac{\tau_2}{\mbbar^3}  
-1.4 \frac{\tau_3}{\mbbar^3} 
-1.6 \frac{\tau_4}{\mbbar^3}
+\epsilon\, \left(0.0002 + 0.0002 \frac{\lbar}{\mbbar}\right) 
\nn\\&\,& \hspace{.1cm} \left.
+ \epsilon^2_{\rm BLM} 0.0005
+ {\left|\frac{V_{ub}}{V_{cb}}\right|}^2 \left(0.12 + 1.3 \frac{\lbar}{\mbbar}\right) 
- \left(0.0031 + 0.006\frac{\lbar}{\mbbar}\right) 
- \left(0.0003 - 0.003 \frac{\lbar}{\mbbar}\right) \right] \nn
\end{eqnarray}
Using the hypothetical data $R^{(4)}_{a} = 2.1558$ and $R^{(4)}_{b} = 0.6788$ we find
\begin{eqnarray}
\lbar = \left(0.64 \pm 0.01_{V_{ub}} \pm 0.01_\epsilon \pm \pm 0.02_{m^3}\right) {\rm GeV}\,, \quad 
\lambda_1 = \left(-0.19 \pm 0.06_{V_{ub}} \pm 0.02_\epsilon \pm \pm 0.02_{m^3}\right) {\rm GeV^2}\,.
\end{eqnarray}

\subsection{Moments with small dependence on $\lbar$ and $\lambda_1$ with no restrictions on $E_\ell$}

\underline{\bf Moment $R[0.8,1,0.1,1.3]$}
\begin{eqnarray}
D_2 &=& R[0.8,1,0.1,1.3] = \hspace{0.1cm}
1.7587\Big[1 
+ 0.01 \frac{\lbar}{\mbbar} 
+ 0.11 \frac{\lbar^2}{\mbbar^2} 
+0.005 \frac{\lambda_1}{\mbbar^2} 
+ 0.97 \frac{\lambda_2}{\mbbar^2} 
+ 0.3 \frac{\lbar^3}{\mbbar^3}
+ 1.2 \frac{\lbar \lambda_1}{\mbbar^3} 
 \\
&\,& \hspace{.1cm}
+ 4.0 \frac{\lbar \lambda_2}{\mbbar^3} 
 - 1.0 \frac{\rho_1}{\mbbar^3} - 
1.7 \frac{\rho_2}{\mbbar^3} 
+0.3 \frac{\tau_1}{\mbbar^3} 
- 1.1 \frac{\tau_2}{\mbbar^3} 
+1.0 \frac{\tau_3}{\mbbar^3} 
+ 1.0 \frac{\tau_4}{\mbbar^3}
-\epsilon\, \left(0.0010  + 0.002 \frac{\lbar}{\mbbar}\right) 
\nn \\
&\,& \hspace{.1cm}\left.
- \epsilon^2_{\rm BLM} 0.0004
 + {\left|\frac{V_{ub}}{V_{cb}}\right|}^2 \left(0.02 -1.2 \frac{\lbar}{\mbbar}\right)  
+ \left(0.0055 + 0.008\frac{\lbar}{\mbbar}\right) 
+ \left(0.0015 + 0.0004 \frac{\lbar}{\mbbar}\right)\right] \nn
\end{eqnarray}
This leads to
\begin{eqnarray}
D_2 = 1.7784 \left[1 \pm 0.0014_{\lambda_1} \pm 0.0016_{\lbar} \pm 0.0007_{V_{ub}}
\pm 0.0004_\epsilon \pm 0.0009_{m^3}\right] 
\end{eqnarray}

\subsection{Moments with small dependence on $\lbar$ and $\lambda_1$, restricted to $E_\ell > 1.5\, {\rm GeV}$}
\underline{\bf Moment $R[0.7,1.6,1.5,1.5]$}
\begin{eqnarray}
D_3 &=& R[0.7,1.6,1.5,1.5] = \hspace{0.1cm}
0.5256\Big[1 
- 0.04 \frac{\lbar}{\mbbar} 
- 0.24 \frac{\lbar^2}{\mbbar^2} 
+0.09 \frac{\lambda_1}{\mbbar^2} 
- 1.16 \frac{\lambda_2}{\mbbar^2} 
- 0.7 \frac{\lbar^3}{\mbbar^3}
- 2.0 \frac{\lbar \lambda_1}{\mbbar^3} 
 \\
&\,& \hspace{.1cm}
- 7.8 \frac{\lbar \lambda_2}{\mbbar^3}  
+ 2.6 \frac{\rho_1}{\mbbar^3} 
+ 2.2 \frac{\rho_2}{\mbbar^3} 
- 0.2 \frac{\tau_1}{\mbbar^3} 
+ 1.3 \frac{\tau_2}{\mbbar^3}
- 1.0 \frac{\tau_3}{\mbbar^3} 
- 1.2 \frac{\tau_4}{\mbbar^3}
+\epsilon\, \left(0.0001  + 0.0001 \frac{\lbar}{\mbbar}\right) 
 \nn \\
&\,& \hspace{.1cm} \left.
+ \epsilon^2_{\rm BLM} 0.00005
- {\left|\frac{V_{ub}}{V_{cb}}\right|}^2 \left(0.13 - 3.0 \frac{\lbar}{\mbbar}\right)  - \left(0.0030 + 0.007\frac{\lbar}{\mbbar}\right) - \left(0.0018 - 0.0008 \frac{\lbar}{\mbbar}\right)\right] \nn
\end{eqnarray}
This leads to
\begin{eqnarray}
D_3 = 0.5166 \left[1 \pm 0.0014_{\lambda_1} \pm 0.0042_{\lbar} \pm 0.0013_{V_{ub}}
\pm 0.00005_\epsilon \pm 0.0012_{m^3}\right] 
\end{eqnarray}
\underline{\bf Moment $R[2.3,1.6,2.9,1.5]$}
\begin{eqnarray}
D_4 &=& R[2.3,1.6,2.9,1.5] =\hspace{0.1cm}
0.6108\Big[1 
- 0.05 \frac{\lbar}{\mbbar} 
- 0.26 \frac{\lbar^2}{\mbbar^2} 
- 0.02 \frac{\lambda_1}{\mbbar^2} 
- 1.17 \frac{\lambda_2}{\mbbar^2} 
- 0.8 \frac{\lbar^3}{\mbbar^3}
- 2.6 \frac{\lbar \lambda_1}{\mbbar^3} 
 \\
&\,& \hspace{.1cm}
- 7.9 \frac{\lbar \lambda_2}{\mbbar^3}  
+ 2.5 \frac{\rho_1}{\mbbar^3} 
+ 1.8 \frac{\rho_2}{\mbbar^3} 
- 0.3 \frac{\tau_1}{\mbbar^3} 
+ 1.0 \frac{\tau_2}{\mbbar^3} 
- 1.0 \frac{\tau_3}{\mbbar^3} 
- 1.2 \frac{\tau_4}{\mbbar^3}
+\epsilon\, \left(0.0001  + 0.0003 \frac{\lbar}{\mbbar}\right)
 \nn \\
&\,& \hspace{.1cm} \left.
+ \epsilon^2_{\rm BLM} 0.0002
. - {\left|\frac{V_{ub}}{V_{cb}}\right|}^2 \left(0.12 - 3.5 \frac{\lbar}{\mbbar}\right)  
- \left(0.0027 + 0.007\frac{\lbar}{\mbbar}\right) 
- \left(0.0013 - 0.003 \frac{\lbar}{\mbbar}\right)\right] \nn
\end{eqnarray}
This leads to
\begin{eqnarray}
D_4 = 0.6016 \left[1 \pm 0.0034_{\lambda_1} \pm 0.0042_{\lbar} \pm 0.0017_{V_{ub}}
\pm 0.0002_\epsilon \pm 0.0011_{m^3}\right] 
\end{eqnarray}
\end{widetext}


\begin{thebibliography}{99}
%
\bibitem{ope}
M.~A.~Shifman and M.~B.~Voloshin,
Sov. \ J.\ Nucl.\ Phys.\ {\bf 41}, 120 (1985);
J.~Chay, H.~Georgi and B.~Grinstein,
Phys.\ Lett. \ B{\bf 247}, 399 (1990).
%
\bibitem{opecorr1}
I.~I.~Bigi \textit{ et al.},Phys.\ Lett.\  B {\bf 293}, 430 (1992)
[(E) \textit{ibid.} B{\bf 297} 477 (1993)];
I.~I.~Bigi,\textit{ et al.}, 
Phys.\ Rev.\ Lett.\  {\bf 71}, 496 (1993).
%
\bibitem{opecorr2}
A.~V.~Manohar and M.~B.~Wise,
Phys.\ Rev.\ D {\bf 49}, 1310 (1994); 
B.~Blok \textit{ et al.}
Phys.\ Rev.\ D {\bf 49}, 3356 (1994)[(E) \textit{ibid.} D{\bf 50} 3572 (1994)];
T.~Mannel,
Nucl.\ Phys. \ B {\bf 413} 396 (1994).
%
\bibitem{wisemanohar}
A. Manohar, M. Wise, Heavy Quark Physics, Cambridge University Press, 2000\,.
%
\bibitem{CKMtalk}
E.~Barberio, {\it Reports from WGI, Workshop on the CKM Unitarity Triangle}, CERN, Feb 13-16, 2002\,.
\bibitem{Isgurduality}
N.~Isgur,
Phys.\ Lett.\ B {\bf 448}, 111 (1999).
\bibitem{leptonmoment2}
M.~Gremm, A.~Kapustin, Z.~Ligeti and M.~B.~Wise,
Phys.\ Rev.\ Lett.\ {\bf 77}, 20 (1996).
%
\bibitem{leptonmoment3}
M.~Gremm and A.~Kapustin,
Phys.\ Rev.\ D {\bf 55}, 6924 (1997).
%
\bibitem{leptonmoment4}
M.~Gremm and I.~Stewart,
Phys.\ Rev.\ D {\bf 55}, 1226 (1997).
%
\bibitem{renormalon}
I.~I.~Bigi, M.~A.~Shifman, N.~G.~Uraltsev and A.~I.~Vainshtein,
Phys.\ Rev.\ D {\bf 50}, 2234 (1994).
M.~Beneke and V.~M.~Braun,
Nucl.\ Phys.\ B {\bf 426}, 301 (1994);
%
\bibitem{luke_renormalon}
M.~E.~Luke, A.~V.~Manohar and M.~J.~Savage,
Phys.\ Rev.\ D {\bf 51}, 4924 (1995).
\bibitem{alphaspec}
M.~Jezabek and J.H.~Kuhn,
Nucl.\ Phys.\ B {\bf 320}, 20 (1989),
A. ~ Czarnecki and M. Jezabek,
Nucl.\ Phys.\ B {\bf 427}, 3 (1994).
%
\bibitem{renorupsilon}
A.~Hoang, Z.~ Ligeti and A.~Manohar,
Phys.\ Rev.\ D {\bf 59}, 074017, (1999).
%
\bibitem{upmassmanohar}
A.~Hoang, Z.~ Ligeti and A.~Manohar,
Phys.\ Rev.\ Lett.\ {\bf 82}, 277, (1999);
A.~H.~Hoang and T.~Teubner,
Phys.\ Rev.\ D {\bf 60}, 114027 (1999)\,.
%
\bibitem{cleoprl}
B.~Barish {\it et al.}  
Phys.\ Rev.\ Lett.\  {\bf 76}, 1570 (1996)\,.
\bibitem{wang_thesis}
Roy Wang, Ph.D. Thesis, University of Minnesota (1994).
%
\bibitem{atwood}
D.~Atwood and W.J. ~Marciano 
Phys. \ Rev. \ D {\bf 41}, 5 (1990)\,.
%
\bibitem{mypapers}
A.~F.~Falk and M.~E.~Luke,
Phys.\ Rev.\ D {\bf 57}, 424 (1998);
C.~W.~Bauer
Phys.\ Rev.\ D {\bf 57}, 5611 (1998),
[Erratum-ibid.\ D {\bf 60}, 099907 (1998)];
C.~W.~Bauer and C.~N.~Burrell,
Phys.\ Lett.\ B {\bf 469}, 248 (1999);
Phys.\ Rev.\ D {\bf 62}, 114028 (2000).

\bibitem{vacuumsat}
I.~I.~Bigi \textit{ et al.},
Phys.\ Rev.\ D {\bf 52}, 196, (1995).
%
\bibitem{hoang}
A.~H.~Hoang,
Phys.\ Rev.\ D {\bf 59}, 014039 (1999);
Phys.\ Rev.\ D {\bf 61}, 034005 (2000);
M.~Beneke and A.~Signer,
Phys.\ Lett.\ B {\bf 471}, 233 (1999);
%
B.~Grinstein and Z.~Ligeti,
Phys.\ Lett.\ B {\bf 526}, 345 (2002).
%
\bibitem{Vub}
C.~W.~Bauer, Z.~Ligeti and M.~E.~Luke,
Phys.\ Lett.\ B {\bf 479}, 395 (2000)\,;
Phys.\ Rev.\ D {\bf 64}, 113004 (2001)\,;
A.~K.~Leibovich, I.~Low and I.~Z.~Rothstein,
Phys.\ Rev.\ D {\bf 61}, 053006 (2000)\,;
Phys.\ Lett.\ B {\bf 486}, 86 (2000)\,;
A.~Bornheim {\it et al.}  
hep-ex/0202019.







\end{thebibliography}
\end{document}